\pdfoutput=1
\documentclass{PoS}

\usepackage{amsmath}
\usepackage{amssymb}
\usepackage{cite}
\usepackage{mathtools}

\newcommand{\nn}{\nonumber}

\newcommand{\bea}{\begin{eqnarray}}
\newcommand{\eea}{\end{eqnarray}}

\usepackage{dcolumn} 
\usepackage{longtable} 
\usepackage{booktabs}
\usepackage{tabularx}

\usepackage{graphicx}
\usepackage{dcolumn}
\usepackage{bm}

\usepackage{epstopdf}
\usepackage{epsfig}
\usepackage{graphics}
\usepackage{xspace}
\usepackage{slashed}
\usepackage{feynmp}
\usepackage{ams fonts}
\usepackage{sidecap}
\usepackage{longtable} 
\usepackage{booktabs}
\usepackage{color}
\usepackage{subfigure}
\usepackage[section]{placeins}
\usepackage{subfigure}

\newcommand{\be}{\begin{equation}}
\newcommand{\ee}{\end{equation}}

\newcommand{\eq}{Eq.~}

\newcommand{\fig}{Fig.~}

\newcommand{\la}{\label}

\newcommand{\ba}{\begin{eqnarray}}
\newcommand{\ea}{\end{eqnarray}}

\title{Glue Spin $S_G$ in The Longitudinally Polarized Nucleon}

\ShortTitle{Glue Spin $S_G$ in The Nucleon}

\author{\speaker{Raza Sabbir Sufian}$^{1,\dagger}$, Michael J. Glatzmaier$^{1,\ddagger}$, Yi-Bo Yang$^{1,\star}$, Keh-Fei Liu$^{1,\circledast}$, Mingyang Sun$^{1}$
\vspace*{-0.5cm}
\begin{center}
\large{
\vspace*{0.4cm}
\includegraphics[scale=0.20]{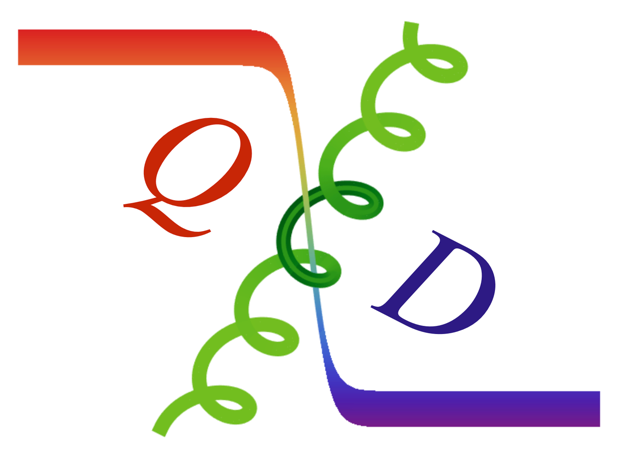}\\
\vspace*{0.4cm}
($\chi$QCD Collaboration)
}
\end{center}
\\
$^{1}$Department of Physics and Astronomy, University of Kentucky, Lexington, KY 40506\\
$^{\dagger}$E-mail: \email{sabbir.sufian@uky.edu}\\
$^{\ddagger}$E-mail: \email{michael.glatzmaier@gmail.com}\\
$^{\star}$E-mail: \email{ybyang@pa.uky.edu}\\
$^{\circledast}$E-mail: \email{liu@pa.uky.edu}
}

\abstract{
\vspace{0.5cm}
We present a lattice QCD calculation of the glue spin $S_G$ in the nucleon for the first time. 
It was recently shown that the first moment of the glue helicity distribution could be obtained through the 
cross-product of the the electric field $\vec{E}$ and the physical gauge field  $\vec{A}_{phys}$ with the 
non-Abelian Coulomb gauge condition, i.e. $\int d^3 x\, \,\vec{E}(x) \times \vec{A}_{phys}(x)$ in the infinite momentum frame. We use the gauge field tensor from the overlap Dirac operator to check the frame dependence and calculate glue spin with several momenta. 
The calculation is carried out with valence overlap fermion on 2+1 flavor DWF gauge configurations on the 
$24^3 \times 64$ lattice with $a^{-1}=1.77$ GeV
with the light sea quark mass corresponding to a pion mass of 330 MeV.}

\FullConference{The 32nd International Symposium on Lattice Field Theory\\
		 23-28 June, 2014\\
		 Columbia University New York, NY}

\begin{document}

\section{Introduction}

The complete decomposition of nucleon spin into its quark and glue components has been a long challenging problem in QCD  since the first discovery of `proton spin crisis'~\cite{emc1}. To find out where the spin of the proton comes from, it is expected that it should compose of the quark spin, quark orbital angular momentum, glue spin, and glue orbital angular momentum in a sum rule
\bea\la{eq1}
{\bf{J}} = {\bf{S}}_q + {\bf{L}}_q + {\bf{S}}_G + {\bf{L}}_G
\eea
There has been a crucial question for a long time as to whether such a decomposition exists with gauge-invariant local operators and, further more, if they can be measured experimentally. In this work we particularly concentrate on the glue spin operator in regard to the feasibility of calculating its matrix element on the lattice.\\
A recent calculation~\cite{deka1} of quark and glue momenta and angular momenta in the proton on a quenched lattice has found quark spin contributes $\sim 25\%$~\cite{Dong:1995rx}, glue angular momentum contributes 
$\sim 28\%$ and quark orbital angular momentum contributes $\sim 47\%$ of proton spin based on the nucleon spin decomposition with the energy-momentum tensor in the Belafonte formalism~\cite{ji1}. In this case, the glue angular momentum cannot be decomposed into spin and orbital angular momentum. The next major question is if it is at all possible to decompose the glue total angular momentum into its orbital and spin components as suggested in the canonical formalism, but in a gauge invariant fashion.\\
A recent analysis~\cite{florian} of high-statistics 2009 {\Large{S}}TAR~\cite{star} and {\Large{P}}HENIX~\cite{phenix} data showed evidence of non-zero glue helicity in the proton. For $Q^2=10$ $GeV^2$, they found gluon helicity distribution $\Delta g(x,Q^2)$ positive and away from zero in the momentum fraction region $0.05\leq x \leq0.2$. However, the result presented in~\cite{florian} has very large uncertainty in the small $x$-region.  \\
In this work, we carry out a lattice calculation of the glue spin in the nucleon based on the theoretical framework in~\cite{ji2}. We shall present our preliminary result on the momentum dependence of the glue spin in the nucleon.

\section{Theoretical Framework}
The total glue helicity which is the integral of the glue helicity distribution, i.e. 
$\Delta G =\int_0^1 \Delta g(x) dx$, is defined as~\cite{manohar},
\bea\la{eq3}
\Delta G &=& \int dx \frac{i}{2xP^+}\int \frac{d\xi^-}{2\pi}e^{-ixP^+\xi^-} \langle PS | F^{+\alpha}_a (\xi^-)\mathcal{L}^{ab}(\xi^-,0)\tilde{F}^+_{\alpha,b}(0) | PS\rangle
\eea
where the light front coordinates are $\xi^{\pm}=(\xi^0\pm\xi^3)/\sqrt{2}$.  The proton plane wave state is written as $|PS\rangle$, with momentum $P^\mu$ and polarization $S^\mu$. The dual gauge field tensor, $\tilde{F}^{\alpha\beta}=\frac{1}{2}\epsilon^{\alpha\beta\mu\nu}F_{\mu\nu}$ and the light cone gauge-link is in the adjoint representation $\mathcal{L}(\xi^-,0)=\mathcal{P}\exp\Bigg[-i g\int^{\xi^-}_0 \mathcal{A}^+(\eta^-,0_\perp)d\eta^-\Bigg]$ with $\mathcal{A}^+\equiv T^cA^+_c$. 
\eq(\ref{eq3}) is gauge invariant but its partonic interpretation is clear only in the light cone gauge.  One cannot evaluate this expression on the lattice due to its dependence on real time, given by $\xi^-$. 
After integrating the longitudinal momentum $x$, the light-cone operator for the matrix element has the following expression~\cite{Hatta:2011zs,ji2} 
\bea\la{eq4}
\tilde{S}_g &=& \Bigg[ \vec{E}^a(0)\times (\vec{A}^a(0)-\frac{1}{\nabla^+} (\vec{\nabla}A^{+,b})\mathcal{L}^{ba}(\xi^-,0))\Bigg]^z
\eea

This expression is analogous to the gauge-invariant glue spin density operator 
\bea\la{ExA}
\vec{S}_g =  \vec{E}^a\times\vec{A}^a_{phys}
\eea
proposed by X. Chen \textit{et al.}~\cite{chen1}, where $A_{\mu\, phys}$ is the physical component of
the gauge field $A_{\mu}$ which is decomposed to $A_{\mu\,phys}$ and a pure gauge part
as in QED,
\bea\la{Amu}
A_{\mu} = A_{\mu\,phys} + A_{\mu\,pure}
\eea
They transform homogeneously and inhomogeneously with respect to gauge transformation respectively,
\bea\label{eq5}
A_{\mu\,phys}  &&  \rightarrow A'_{\mu\,phys}= g A_{\mu\,phys}g^{-1} \nn \\
A_{\mu\,pure} &&  \rightarrow A'_{\mu\,pure} = gA_{\mu\,pure}g^{-1}+\frac{i}{g_0}g\partial_\mu g^{-1}, \nn \\
\eea
where $g$ is the gauge transformation matrix and $g_0$ is the coupling constant. In oder to have a unique solution, conditions are set as follows: the pure gauge part does not give rise to a field tensor by itself and $A_{\mu\,phys}$ satisfies the non-Abelian Coulomb gauge condition
\bea
F_{\mu\nu\,pure} &&=\partial_\mu A_{\nu\,pure} -\partial_{\nu}A_{\mu\,pure}+ig_0[A_{\mu\,pure},\, 
A_{\nu\,pure}]=0\nn \\
D_i\, A_{i\,phys} &&= \partial_i\, A_{i\,phys}-ig_0[A_i,\,A_{i\,phys}] =0. 
\eea
It is shown by Ji, Zhang, and Zhao~\cite{ji2} that when boosting the glue spin density operator $\vec{S}_g$ in Eq.(\ref{ExA}) to the infinite momentum frame (IMF), the second term in the parentheses on the right side of
Eq. (\ref{eq4}) is $\vec{A}_{pure}$. Thus $\tilde{S}_g$ is just the glue spin density operator $\vec{S}_g$ in
the IMF along the direction of the moving frame.\\ 
It can be shown~\cite{yl14} that $A_{\mu\,phys} = g_C^{-1} (A_C)_\mu g_C$ where $A_C$ is the gauge potential
fixed to the Coulomb gauge and $g_C$ is the gauge transformation that fixes the Coulomb gauge.
Since $\vec{S}_g$ is traced over color, spin operator is then
\bea\label{eq7}
\vec{S}_G = \int d^3 x\, Tr\, (g_C\vec{E}g_C^{-1}\times\vec{A}_C) =  \int d^3 x\, Tr\, (\vec{E}_C\times
\vec{A}_C)
 \eea
where $\vec{E}_C$ is the electric field in the Coulomb gauge. Although it is gauge invariant since both
$E$ and $A_{phys}$ transform homoeneouly, it is frame dependent and depends on the proton momentum.
Its IMF value corresponds to $\Delta G$ as measured experimentally from high energy proton-proton
scattering. The important outcome of the derivation is that glue spin content is amenable to lattice QCD
calculation. To the extent that it can be calculated at large enough momentum frame of the proton with
enough precision, it can be compare to the experimental glue helicity $\Delta G$.

\section{Lattice Calculation}

\subsection{Simulation Details}
We use valance overlap fermion on $(2+1)$ flavor RBC/UKQCD DWF gauge configurations on the$24^3\times64$ lattice~\cite{rbc}. The number of configurations we use in our calculation is 203. The corresponding sea quark masses in the lattice units are $am_l=0.005$ and $am_s=0.04$. The inverse lattice spacing is $a^{-1}=1.77$ GeV and $L=2.7$ fm. The light sea mass corresponds to a pion mass of 330 MeV.\\
The nucleon propagator is constructed from the grid-$8$ smeared source with $Z_3$ noise using low mode 
substitution~\cite{anyi,ming}. We set the source time slices at $t=0$ and $t=32$ and then shift the sources from $t=0$ to $t=31$ and $t=32$ to $t=63$ in pairs to increase statistics.\\
We use three different momenta $p =1,2$ and 3 where $p$ is defined as
\bea
p=\frac{2\pi n}{La} , \text{ for n=0,1,2,...}
\eea
\subsection{Construction of $F_{\mu\nu}$ From The Overlap Dirac Operator $(D^{ov})$ }

It has been shown in \cite{liu1} that, on a hyper-cubic lattice with lattice spacing $a$, the gauge field strength tensor $F_{\mu\nu}(x)$ can be obtained from $D^{ov}$ operator
\bea
tr_s\, \sigma_{\mu\nu}\, D^{ov}(x,\,x) = c^T(\rho)\, a^2\, F_{\mu\nu}(x) + \mathcal{O}(a^3),
\eea
where $tr_s$ denotes the spinor trace, $D^{ov}(x,\,x)$ is the diagonal matrix element of the overlap operator 
at $(x,\,x)$. The proportional constant $c^T = c^T(\rho)$ is a function of the overlap mass parameter $\rho$ and $\sigma_{\mu\nu} = \frac{1}{2i}[\gamma_\mu,\gamma_\nu]$. It has been found that in the study of the QCD vacuum structure with the topological charge density derived from $D^{ov}$ produces better signal than those obtained from gauge links~\cite{ivan1,ivan2}. This is also true when the gauge field tensor from the overlap operator is used to calculate the glue momentum and angular momentum in the nucleon~\cite{deka1}.
This is because the non-ultralocal beahvior of $D^{ov}$ serves as an efficient filter of UV fluctuations through chiral smearing. Therefore, in this work we construct $F_{\mu\nu}(x)$ from the $D^{ov}$ operator and obtain the electric field from $E_i =-F_{4i}$.

\subsection{Construction of Coulomb gauge fixed potential $(A_C)_\mu(x)$}
The standard way of gauge fixing on the lattice \cite{gfix} is equivalent to maximizing the gauge functional
\bea
F_g[U]= \mathcal{R}e\sum_{\mu,x} tr[U^g_\mu(x)+U^g_\mu(x-\hat{\mu})^\dagger]
\eea
with respect to gauge transformations $g(x) \in SU(3)$ where
\bea
U^g_{\mu}(x) \equiv g(x)U_\mu(x)g(x+\hat{\mu})^\dagger.
\eea
We have defined the Coulomb gauge-fixed $A_\mu(x)$ to be,
\bea
(A_C)_\mu(x) \equiv \Bigg[\frac{U_\mu(x)-U^\dagger_\mu(x)}{2iag_0}\Bigg]_{traceless}
\eea
where $U_\mu(x)=e^{iag_0A_\mu(x)}$. \\

We calculate tadpole improved lattice gauge coulping $\tilde{g}_0$ for Iwasaki gauge action and therefore tadpole improve the glue spin density operator $(\vec{E_C}\times\vec{A_C})$ in our calculation based on mean field improvement in~\cite{lepage}. To obtain tadpole improved $\vec{E_C}$ calculated from $D^{ov}$ operator, we use tadpole improved value of $c^T(\rho)$.

\subsection{Disconnected Insertion of The Three Point Correlation Function}

In order to extract the glue spin contributions to the nucleon, we compute the ratio of the disconnected three-point function to the two-point function with the nucleon propagator source and sink located at $t_0$ and $t$. We insert the  glue spin value, $L_i(t')=\int d^3 x\, (\vec{E_C}\times\vec{A_C})(x,\,t')$ at some insertion time $t'$. Then the ratio of disconnected insertion three-point function to two-point function can be written as
\bea
\tilde{R}(t,t',t_0) = \frac{\Bigg\langle S_N(t,t_0)(L(t')-\langle L(t') \rangle )\Bigg\rangle}{\langle S_N(t,t_0)\rangle}
\eea
where $S_N(t,t_0)$ is the nucleon propagator. We improve the statistics by using the summed ratio method \cite{sum} and denote the summed ratio by $R'(t,t_0) = \sum_{t'=t_0+1}^{t-1} \tilde{R}(t,t',t_0)$, where $\tilde{R}(t,t',t_0)$ is summed over $t'$ between $t_0+1$ and $t-1$ inclusive. $R'(t,t_0)$ has a linear behavior in the region where $t$ is assumed to be large enough and therefore we label $R'(t,t_0)$ as $R(t)$ in \fig\ref{fig:1} and \fig\ref{fig:2}. The slope of the linear region of the $R(t)$ plot gives a measure of the glue spin in the nucleon for a given momentum of the nucleon. 
 
 \section{Numerical Results} 
 We use two different masses for the valance quarks, namely $m_q=0.0203$ and $m_q=0.0756$ in lattice units which correspond to $m_\pi\simeq 380$ MeV and $m_\pi\simeq 640$ MeV respectively. \\

 \begin{figure}
\begin{center}
    \includegraphics[width=.42\textwidth]{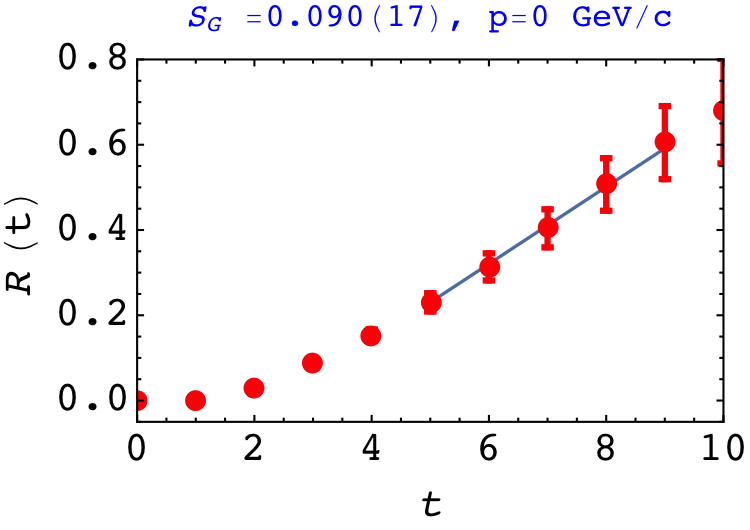}
    \includegraphics[width=.42\textwidth]{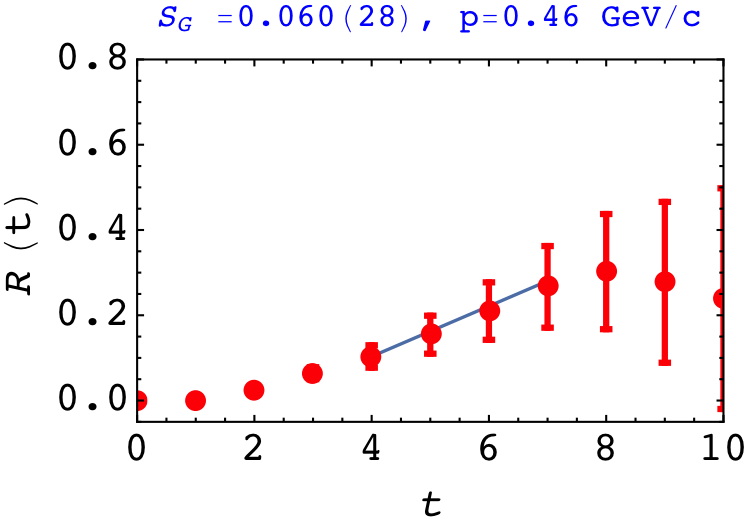}
     \includegraphics[width=.42\textwidth]{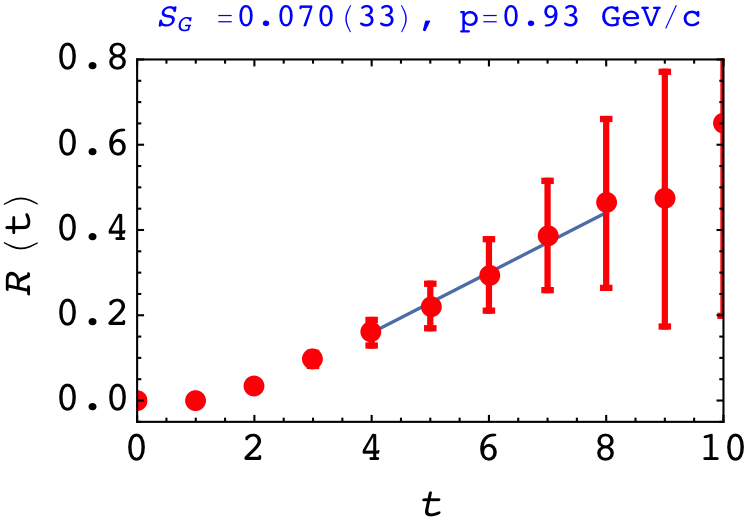}
\end{center}
\caption{Summed ratio $R(t)$ for $m_\pi=640\, MeV$.}
\la{fig:1}
\end{figure}
\fig\ref{fig:1} shows our signal from the constructed values of $R(t)$ for the quark mass equivalent to $m_\pi\simeq 640$ MeV. In our current analysis the heavier quark mass produces better signal, especially for momenta $p=1\,(0.46\,\,GeV/c)$ and $p=2\,(0.93\,\,GeV/c)$ when we fit the slope to get an estimate of $S_G$. 
 \begin{figure}
\begin{center}
    \includegraphics[width=.80\textwidth]{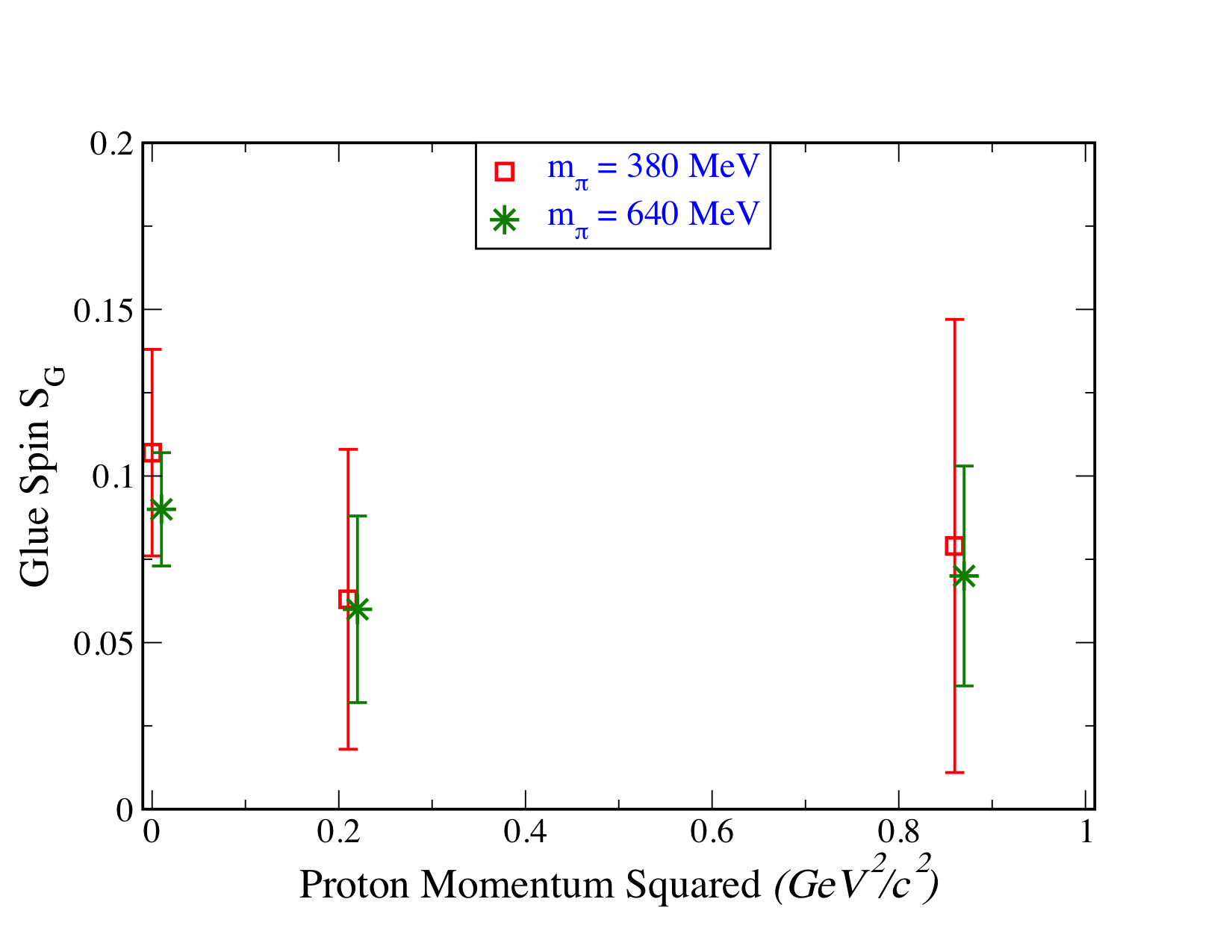}
\end{center}
\caption{Fit results of glue spin $(S_G)$ in longitudinally polarized proton.}
\la{fig:2}
\end{figure} 
\fig\ref{fig:2} shows the fit results of $S_G$ for different quark masses and momenta. At this stage due to the large errors in the fit results of $p=0.46\,\,GeV/c$ and $p=0.93\,\,GeV/c$ momenta of the proton, we cannot extrapolate to any value of $S_G$ in the large momentum frame of the proton. However it is clear that our analysis shows a positive spin $S_G$ for
proton at rest.  Fro the case of $p\,a=0$ and $m_\pi = 640$ MeV, the value of $S_G$ is ${\bf{(18\pm3.4)\%}}$ of the total proton spin of $\frac{1}{2}$.


Lattice perturbation calculation is being carried out~\cite{mike} to match this lattice result to that of the continuum. 
Non-perturbative renormalization based on momentum and angular momentum sum rules~\cite{deka1} is also being pursued. 
\section{Conclusion}

Since the errors of our present calculation are large for $p \neq 0$  (as seen in \fig\ref{fig:2}), we cannot extrapolate $S_G$ to the large momentum frame of the proton and draw any conclusion of the momentum dependence of  $S_G$. The most important result one can conclude from the fitted values of $S_G$ is that the contribution of glue spin to nucleon is non-zero and positive for the proton at rest.  

We plan to increase our statistics of the nucleon propagator and improve signal of the glue spin density operator and perform analysis with larger momenta. We also aim to perform similar analysis on the $32^3\times 64$ lattice at a smaller lattice spacing.\\

K.F. Liu would like to thank Y. Hatta, X. Ji and Y. Zhao for useful discussions.

\end{document}